# Model Multiplicity (UML) Versus Model Singularity in System Requirements and Design

Sabah Al-Fedaghi
*sabah.alfedaghi@ku.edu.kw, salfedaghi@yahoo.com*
Computer Engineering Department, Kuwait University, Kuwait

**Summary**
A conceptual model can be used to manage complexity in both the design and implementation phases of the system development life cycle. Such a model requires a firm grasp of the abstract principles on which a system is based, as well as an understanding of the high-level nature of the representation of entities and processes. In this context, models can have distinct architectural characteristics. This paper discusses model multiplicity (e.g., unified modeling language [UML]), model singularity (e.g., object-process methodology [OPM], thinging machine [TM]), and a heterogeneous model that involves multiplicity and singularity. The basic idea of model multiplicity is that it is not possible to present all views in a single representation, so a number of models are used, with each model representing a different view. The model singularity approach uses only a single unified model that assimilates its subsystems into one system. This paper is concerned with current approaches, especially in software engineering texts, where multimodal UML is introduced as *the* general-purpose modeling language (i.e., UML is modeling). In such a situation, we suggest raising the issue of multiplicity versus singularity in modeling. This would foster a basic appreciation of the UML advantages and difficulties that may be faced during modeling, especially in the educational setting. Furthermore, we advocate the claim that a multiplicity of views does not necessitate a multiplicity of models. The model singularity approach can represent multiple views (static, behavior) without resorting to a collection of multiple models with various notations. We present an example of such a model where the static representation is developed first. Then, the dynamic view and behavioral representations are built by incorporating a decomposition strategy interleaved with the notion of time.

## 1. Introduction

A conceptual model is a way of conceptualizing (depicting or imitating) how entities and processes in a certain part of the world (e.g., physical, social) work. For the model to be developed, the original world phenomenon must be projected (counterparts developed) in the abstract domain to match these entities and processes on the basis of common or theoretical conceptualization. Here, a phenomenon refers to the stable and general features of a system of interest, where a system is "a collection of elements, related to one another, exhibiting a collective behavior" [1]. Conceptual models are used to support the design of software, business processes, enterprise documentation, etc. [2]. In this context, modeling requires grasping the abstract principles on which the system is based, as well as understanding the high-level nature of the representation of entities and processes.

In the requirements elicitation phase of development, a conceptual model involves collecting information for the purpose of building a representation of the targeted system. According to [3], "No other part of the conceptual work is as difficult as establishing the detailed technical requirements… No other part of the work so cripples the resulting system if done wrong. No other part is as difficult to rectify later." A requirements engineering-based conceptualization [4] can be applied to manage this involved complexity [5].

Requirements engineering begins with interaction among stakeholders (e.g., managers, customers, software engineers, and end users) and continues during the modeling process, where needs are specified, scenarios are described, functions and features are delineated, and project constraints are identified [6]. The first task in such a process is the requirements inception (gathering), which is concerned with the objectives for the system, what is to be accomplished, and how the system is to be used on a day-to-day basis [6].

### 1.1 Styles of Modeling

To understand the process of creating conceptual models, one must understand that several styles of modeling exist:
- Model multiplicity (e.g., unified modeling language [UML]),
- Model singularity (e.g., object-process methodology [OPM], thinging machine [TM]), or
- Heterogeneous model [12]

The basic idea of model multiplicity is that it is not possible to present all views in a single representation, so a number of models are used, with each model representing an alternative view. The model singularity approach uses only one integrated model. Each of these styles of modeling has its own strengths and weaknesses.

## 1.2 UML as the Standard Modeling Language

In our computer engineering department, the classical text *Software Engineering: A Practitioner's Approach* by Pressman and Maxim [6] is used in the Introduction to Software Engineering course. The book focuses on a case study called SafeHome to teach how to conduct such an initial phase of development using UML. The modeling concentrates solely on UML and its 14 diagrams in accomplishing such a task, as UML is the go-to option for explaining software design models [7]. According to Oliver [7], "What makes UML well-suited to and much-needed for software development is its flexibility. UML is a rich and extensive language that can be used to model not just object-oriented software engineering, but application structure, behavior, and business processes too." Sharma et al. [5] observed that a UML conceptual model underlines three major elements: building blocks (e.g., things, relationships, and diagrams), rules (e.g., names, scopes, and execution), and common mechanisms (e.g., specifications, stereo types, and tagged values).

However, UML has grown in complexity, which makes many people feel as though they are better off without it [7]. Complexity is the number-one problem in the software industry [8]. It is common for students to have difficulty with absorbing UML due to the involved complications [9]. Often, students think that UML diagrams are useless and serve only as documentation that no one reads [10]. These difficulties in UML originate from the multiplicity of the model paradigm used in existing object-oriented system analysis methods for specifying various system aspects [11].

## 1.3. Problem of Concern

When it comes to teaching an introductory course in software engineering modeling, almost all current texts treat the multimodal UML methodology as general-purpose modeling, which gives students the impression that UML *is* modeling and that modeling *is* UML. This one-sided picture completely ignores the difficulty of multiplicity, as well as alternative modeling methodologies that include singularity or mixture models. In the educational environment, both modeling as a separate topic and various modeling styles should at least be mentioned before UML is adopted as the selected modeling language.

To illustrate the types of claims involved, we consider the following statements, which are common in textbooks: "just as building architects create blueprints for a construction company to use, software architects create UML diagrams to help software developers to build software" [6], and "if you understand UML, you can much more easily understand and specify a system and explain the design of that system to others" [6].

In this paper, we propose introducing modeling and highlighting the issue of model multiplicity versus model singularity before concentrating exclusively on UML. This would give students a basic understanding of the UML advantages and difficulties that they may face during their modeling exercises.

The next section provides further explanation about multiplicity versus singularity in modeling. For the sake of making this paper self-contained, section 3 summarizes our main tool for analyzing modeling, a TM. Section 4 introduces the foundation of modeling in a TM.

## 2. Multiplicity vs. Singularity

A conceptual view is a representation of a system from the perspective of a related concern (e.g., structural aspect, behavioral aspect, functional aspect, logical aspect, organizational aspect, infrastructural aspect). It is a piece of the model that is still small enough to be comprehended and that also contains relevant information about a particular concern [13]. According to the Institute of Electrical and Electronics Engineers 1471 [14] standard for architecture modeling, a view is a depiction of a whole system from the perspective of related concerns. Whittle et al. [15] stated that an aspect of model composition is the special case of the more general problem of the fusion of models in one model, or the presence of a crosscut base model. Kruchten [16] defined "4+1" views as those representing different viewpoints: the logical view, process view, physical view, development architecture view, and use case view. The last view is a holistic view that reflects the process associated with a set of system requirements. Tension in the singularity/multiplicity modeling framework provides an approach for dealing with the inherent complexity of systems.

According to [12], the model multiplicity approach utilizes a distinct model for each view. A model is the union of all its representations—or a union of all its *views* [13]. Comprehending a system requires concurrent references to the various models, as well as the creation of abstract associations that link them together. According to Egyed [13], n views need $n(n-1)/2$ (i.e., $O(n^2)$ complexity) ways of integration to be fully integrated. Rather than building the model into an integrating method containing all representations, the alternative is to place this responsibility on the shoulders of the developers.

The model singularity approach produces a single model that enables system specification that assimilates the subsystems into a whole. By contrast, UML unified standards, processes, and views as mostly a segregated collection of subsystems. UML does not determine the semantic integrity needed for necessary qualities such as consistency and completeness.

For example, when one thinks about processes in parallel with objects, the OPM delineates a model singularity framework that avoids the source of this complexity problem through a single integrated model [12]. The origin of the model singularity approach is related to the notion of holistic modeling involving a unified specification that captures the structural, behavioral, and dynamic aspects of the system of interest. Peleg and Dori [11] questioned whether multiplicity/singularity alternative approaches yield a specification that is easier to comprehend.

Some attempts at achieving a heterogeneous style of modeling have been made. For example, Keng-Pei Lin et al. [12] proposed "an approach to enrich UML from model multiplicity to model singularity by creating its kernel model with the structure-behavior coalescence process algebra… Both the UML structure models and behavior models can be derived from this kernel model." Wang [17] proposed the integration and combinational usage of existing modeling languages (i.e., the OPM, UML).

## 3. TM Modeling

The TM model involves a single diagrammatic representation. It articulates the ontology of the world in terms of an entity that is simultaneously a thing and a machine, called a "thimac" [18-27]. A thimac is like a double-sided coin. One side of the coin exhibits the characterizations that the thimac assumes, whereas on the other side, operational processes emerge that provide dynamics. A thing is subjected to doing, and a machine does. We claim that just as the object in object-oriented models is the smallest stand-alone component [13], the thimac is the smallest stand-alone component in TM. However, for simplicity's sake, the thimac is represented in terms of its machine.

The TM notion of a thing is much wider than the notion of an object in object-oriented modeling. The object is originally contrasted with the term "subject." A subject is an actor, and an object is a thing that receives the act. In philosophy, an "object" may be considered to be just "a name for stuff of any kind at any scale" [28]. This is a thing in TM. For example, "John is happy" is about two things: John and happiness. Happiness flows into John. Thimacs are a way for generic constructs to be applied in conceptual modeling to describe the structure/behavior of a world of systems (thimacs). The generic actions in the machine (see Fig. 1) can be described as follows:

**Arrive:** A thing moves toward a machine.
**Accept:** A thing enters the machine. For simplification, we assume that all arriving things are accepted; hence, we can combine the *arrive* and *accept* stages into one stage: the receive stage.
**Release:** A thing is ready for transfer outside of the machine.
**Process:** A thing is changed, but no new thing results.
**Create:** A new thing is born in the machine.
**Transfer:** A thing is input into or output from a machine.

Having a five-action TM machine, or seven when **transfer** includes input and output and **receiv**e includes arrival and acceptance, can greatly reduce the complexity of modeling. After all, the human mind can usually only handle seven distinct things (plus or minus two) at the same time [13]. The underlying cause of complexity is not the number of details "but the number of details of which we have to be aware at the same time" [29]. Additionally, the TM model includes storage and triggering (denoted by a dashed arrow in this study's figures), which initiates a flow from one machine to another. Multiple machines can interact with one another through the movement of things or through triggering. Triggering is a transformation from one series of movements to another.

## 4. A Foundation for Modeling

In this paper, we claim that it is possible to present all views using a single model. We present an example of a model where a static view is developed first, followed by a dynamic view, which is then followed by a behavioral view. The last two views are built upon the static model by using decomposition and inserting the timing element. In this context, a conceptual model is seen as the projection of diverse aspects of concern in the world into a coherent whole that serves as a complete framework for both entire and local aspects. The coherent whole is constituted by means of TMs that include generic things and actions (thimacs) representing components of features of the world.

Accordingly, here, the concern is with definitional features that define the system's internality, which include structural and behavioral aspects. The involved features are fundamental for any system and form a single explanation in the singular model of a coherent system. The singular model is regarded as a coded system that is part of the ontology of the world. Here, "coded" means to be put in a tangible representation, especially symbolic, linguistic, and diagrammatic representation. "Uncoded" means not having to be put in such a representation. In this situation, we can view the model as a coded system. We can say that a "part-of-reality" uncoded system exhibits a structure and behavior, but it has not been recognized as an independent whole. Even in social systems, we find such a phenomenon when we do not recognize some encompassing system that engulfs the existing structure and behavior.

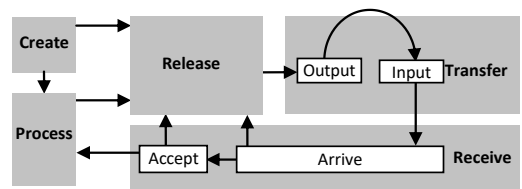

**Fig. 1. The thinging machine.**

For example, the second author of [27], a network engineer working in an actual environment, was surprised to see that the extent of her job included multiple switches, routers, servers, security elements, users, a client computer, protocols, and many other processes when entities and processes were explicitly placed in one unified TM packet, as shown in Fig. 2. Even though she had been practicing her work for years and the "real" uncoded system was right in front of her, she never explicitly thought of it as an independent thing.

The system/model thesis can be related to the Platonic form. In this context, the form is the thing's configuration, in contradistinction to the matter of the thing of which it is composed (*Encyclopedia Britannica*). The forms are typically described as perfect archetypes of which objects in the everyday world are imperfect copies. This means that forms have "beings" as aspects of things in reality. Whether these forms are perfect or imperfect is not an issue in this context. All reality is capable of being expressed as a complex system coded as forms (i.e., their models). Thus, "conceptual models" refer to the coding forms of real systems. "Conceptual" is used because no other creature performs this phenomenon except for a human being who creates, processes, releases, transfers, and receives concepts in a coded visible shape. In a TM, we develop forms of systems as thimacs, which include static (thing) and dynamic (machine) features in agreement with the Hegelian notion that "being" (a real system) is not a static concept. In a TM, a thimac is static and dynamic simultaneously.

Simply stated, models of the world or parts of the world are representations of uncoded systems in the world. Any "existence" (e.g., phenomenon) in reality is accompanied by its system with a structure (e.g., boundary) and behavior. Some models precede their reality (fictitious systems—not of concern here), and some do not. Uncoded subsystems are similarly uncoded parts of reality; nevertheless, they cannot completely replace the whole uncoded system.

A model multiplicity should be accompanied by a model singularity because that is how systems work in reality. The singularity gives an underlying unity to the multiplicity of the parts. In terms of software engineering, such a claim leads to the conclusion that UML requires a 15th model that represents the totality of the uncoded system in such a way that the other 14 models align with this 15th model. Currently, the necessary wholeness is represented, partially and disjointedly, by such diagrams as the class diagram (staticity) and the state diagram (behavior). The state diagram, which is restricted to a single class only, is inaccurately claimed to model behavior. However, one wonders about such a method that represents a behavior without incorporating time explicitly. Here, there is a mix between logical order of actions (or a set of actions) and chronology of events in time. This is applied to activity diagrams, which are a generalization of state diagrams in that "they can also be used to depict *events* or other 'transitional' elements" [13] (italics added).

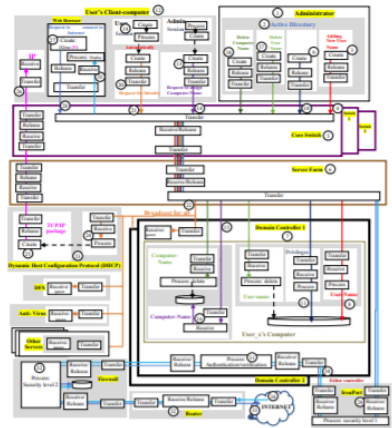

**Fig. 2 Undocumented (and not separately recognized) unified episode of a portion of the job of a network engineer (Adopted from [27]).**

The model singularity approach solves the alignment problem by constructing a view-independent representation of the whole model with views to be derived from such a model. We could define a consistency and completeness rule based on this view-independent representation [13]. According to Egyed 13], "all consistency and completeness rules needed only to be represented in one type of style (language, etc.) and not in a view-dependent form." This approach implies that the model is more than the sum of its views [13]. The stakeholders can then derive views from that model, reconcile the changes with the model, and "all information about a software system is captured with as little redundancy as possible in the model even though the views, which are derived from that model, may repeatedly use the same information and, thus, have redundancy" [13].

Can a TM become the 15th UML diagram? In this situation, some, if not many, UM diagrams (e.g., activity, sequence, and maybe state diagrams) become obsolete. In this case, a view model is developed (coded) from the whole model. Accordingly, this facilitates a focus on how to consider unity and multiplicity within the same modeling system and, more precisely, how to align multiplicity within a singular model. We could suggest that such required unification can be found in the TM and in its generic action that forms generic events. The solution might involve "a community of models" within a whole that are ontologically correlated to one another while being distinct from one another in terms of their purposes. In this case, the multiplicity is subordinate to wholeness. Generally, the modeling is to be understood as a single phenomenon represented by a single holistic model and multiple views.

## 5. Modeling Project

Pressman and Maxim [6] presented a project called the SafeHome project, a home security project that would protect against and/or recognize a variety of undesirable "situations," such as illegal entry, fire, flooding, carbon monoxide levels, and others. It uses wireless sensors to detect each situation, and the homeowner can program it. In addition, it will automatically telephone a monitoring agency when a situation is detected. The purpose is to learn about the principles, concepts, and methods that are used to create requirements and design models. In the preparation process for modeling the required system, lists for the following things are prepared:
- Objects that are part of the environment, produced by the system, and used by the system to perform its functions
- Services (processes or functions) that manipulate or interact with the objects
- Constraints (e.g., cost, size, business rules) and performance criteria (e.g., speed, accuracy).

Ideally, each listed entry should be capable of being manipulated separately. Then, a combined list is created by eliminating redundant entries and adding any new entries that crop up. A mini-spec for the SafeHome object control panel is developed (see Fig. 3). Nonfunctional concerns (e.g., accuracy, data accessibility, security) are registered as nonfunctional requirements. Accordingly, a set of scenarios identify a thread of usage for the system to be constructed, according to Pressman and Maxim [6]. The basic use case (see Fig. 4) for system activation is as follows [6]:

1. The homeowner observes the SafeHome control panel to determine if the system is ready for input. If the system is not ready, a not-ready message is displayed on the LCD display, and the homeowner must physically close windows or doors so that the not-ready message disappears. (A not-ready message implies that a sensor is open, for example, that a door or window is open.)
2. The homeowner uses the keypad to key in a four-digit password. The password is compared with the valid password stored in the system. If the password is incorrect, the control panel will beep once and reset itself for additional input. If the password is correct, the control panel awaits further action.
3. The homeowner selects and keys in "stay" or "away" (see Fig. 3 again) to activate the system. "Stay" activates only perimeter sensors (inside motion-detecting sensors are deactivated). "Away" activates all sensors.
4. When activation occurs, a red alarm light can be observed by the homeowner.

Then, Pressman and Maxim [6] talked about use cases (see Fig. 4) with exceptions that are further elaborated to provide considerably more detail about the interaction. According to Pressman and Maxim [6], use cases for other homeowner interactions would be developed in a similar manner. After this, class-based elements are discussed, with the example of a sensor class (Fig. 5). This is followed by discussing behavioral elements using a state diagram for the software embedded within the SafeHome control panel that is responsible for reading user input. Note the peculiar situation when trying to explain to the students, at this point of modeling, the state diagram that models software. The stream of diagrams continues, thus advocating the greater use of a case diagram (Fig. 6), a collaboration diagram, a sequence diagram, and DFD data models.

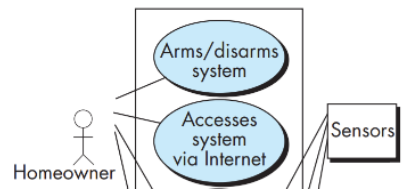

**Fig. 4 Use cases (Incomplete, from [6]).**

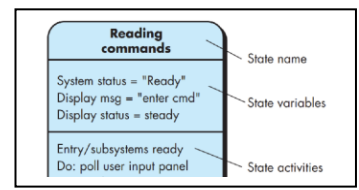

**Fig. 5 State diagram for the software embedded within the SafeHome control panel that is responsible for reading user input (Incomplete, from [6]).**

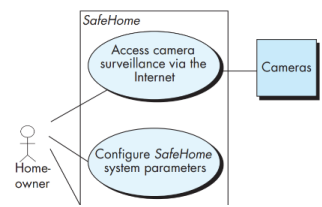

**Fig. 6 Preliminary use case diagram for the SafeHome system (Incomplete, from Pressman and Maxim [6]).**

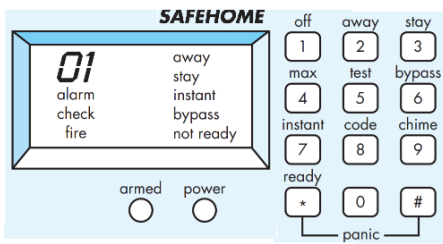

**Fig. 3 Control panel (Rearranged, from [6]).**

The UML swimlane diagram (activity diagram; see Fig. 7) is presented to facilitate the accessing of camera surveillance via the Internet-display camera view function. More diagrams are presented, including a class diagram for the system class, a class diagram for a floor plan, a composite aggregate class diagram, a physical multiplicity diagram (e.g., wall, door, window), a dependency diagram (e.g., camera to display), a package diagram, a state diagram for the control panel class, a sequence diagram (partial) for the SafeHome security function, a sequence diagram for the ActuatorSensor pattern, a class diagram for the ActuatorSensor pattern, etc.

## 6. TM Modeling

A TM focuses on a single (possible multilevel) diagram with in-zooming and out-zooming, if needed, to model the SafeHome. With additional details, the SafeHome model will keep growing as additional parts of the TM diagram are gradually constructed to increasingly extend the model. Pressman and Maxim's [6] UML model would provide more and more details about the SafeHome project to expand the initial TM modeling. We start with the scenario of a homeowner using the keypad mentioned in the previous section.

TM modeling involves two levels: staticity and dynamics. The static model involves spatiality and actionality (generic actions). Spatiality involves recognizing the top thimac areas that partition the model. To draw considerations, we take into account the connections (flows and triggering) among these areas, as shown in Fig. 8. Fig. 8 shows the spatiality of the SafeHome as what must be done first in TM modeling.

The figure contains the following thimacs: screen, keypad, "stay," "away," "beep," sensor regions, and comparison process. The connections are identified from the English description of the SafeHome. For example, "stay" triggers two beeps; hence, we draw a line between "stay" and "beep," and we allocate "stay" close to "beep" to avoid line intersections. Fig. 9 shows further refinement by introducing the type of linking among areas of top thimacs.

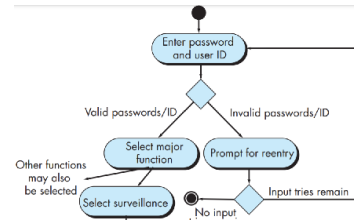

**Fig. 7 Activity diagram for accessing camera surveillance via the Internet-display camera view function (Incomplete, from [6]).**

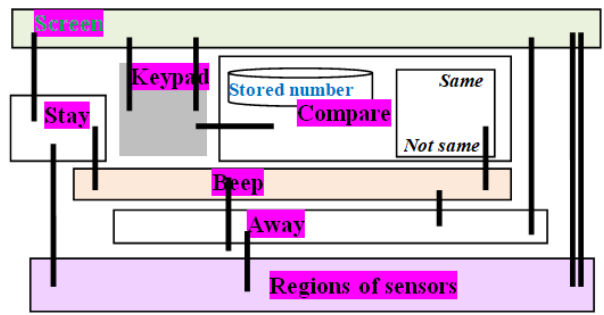

**Fig. 8 Spatiality in the scenario of the homeowner using the keypad.**

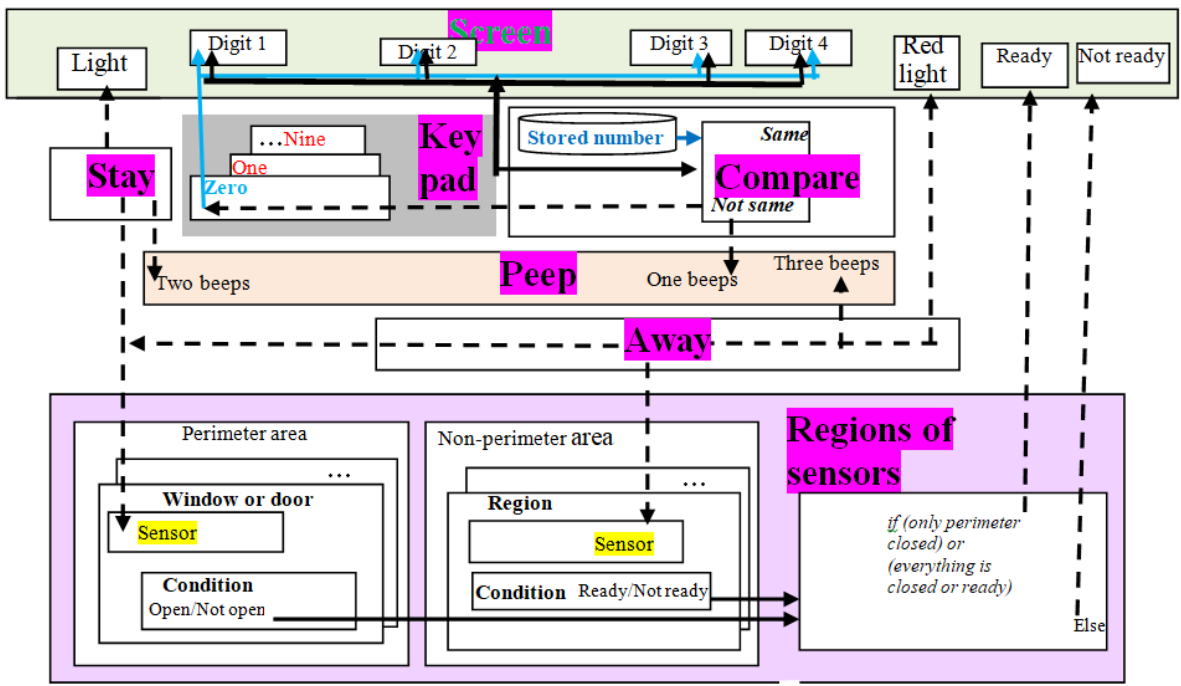

**Fig. 9 The second version of modeling that introduces flows and triggering.**

Finally, Fig. 10 shows the static TM model of the scenario of the homeowner using the control panel. In Fig. 10, the homeowner observes the SafeHome control panel to determine if the system is ready for input (circles 1 and 2 in the lower-right corner of the figure).

"Ready" and "not ready" are triggered by data coming from perimeter and non-perimeter areas (3 and 4). If the condition of a perimeter region (door, window) or a non-perimeter region is not closed or ready, then such data are sent (7 and 8) to where they are processed to trigger "ready" or "not ready" on the screen (1 and 2). The homeowner uses the keypad (9) to key in a four-digit password that flows to the screen (10) to be displayed (11, 12, 13, and 14).

Additionally, the digits flow to a procedure (15) so that they are converted (16) into a number that is compared (17) with a stored number (18). If the password is incorrect (20), the control panel will beep once (21) and reset the digits with zeros (22). If the password is correct, the control panel awaits further action. The homeowner selects and keys in "stay" (23), which makes the state of the control panel "on" (24). This triggers the activation of only perimeter sensors; inside motion-detecting sensors are deactivated (25). The control panel beeps twice (26), and a stay light is lit (27). If the homeowner selects and keys in "away," then this makes the state of the control panel "on" (28). This triggers the activation of all sensors (25 and 29). The control panel beeps three times (30), and the homeowner can observe a red alarm light (31).

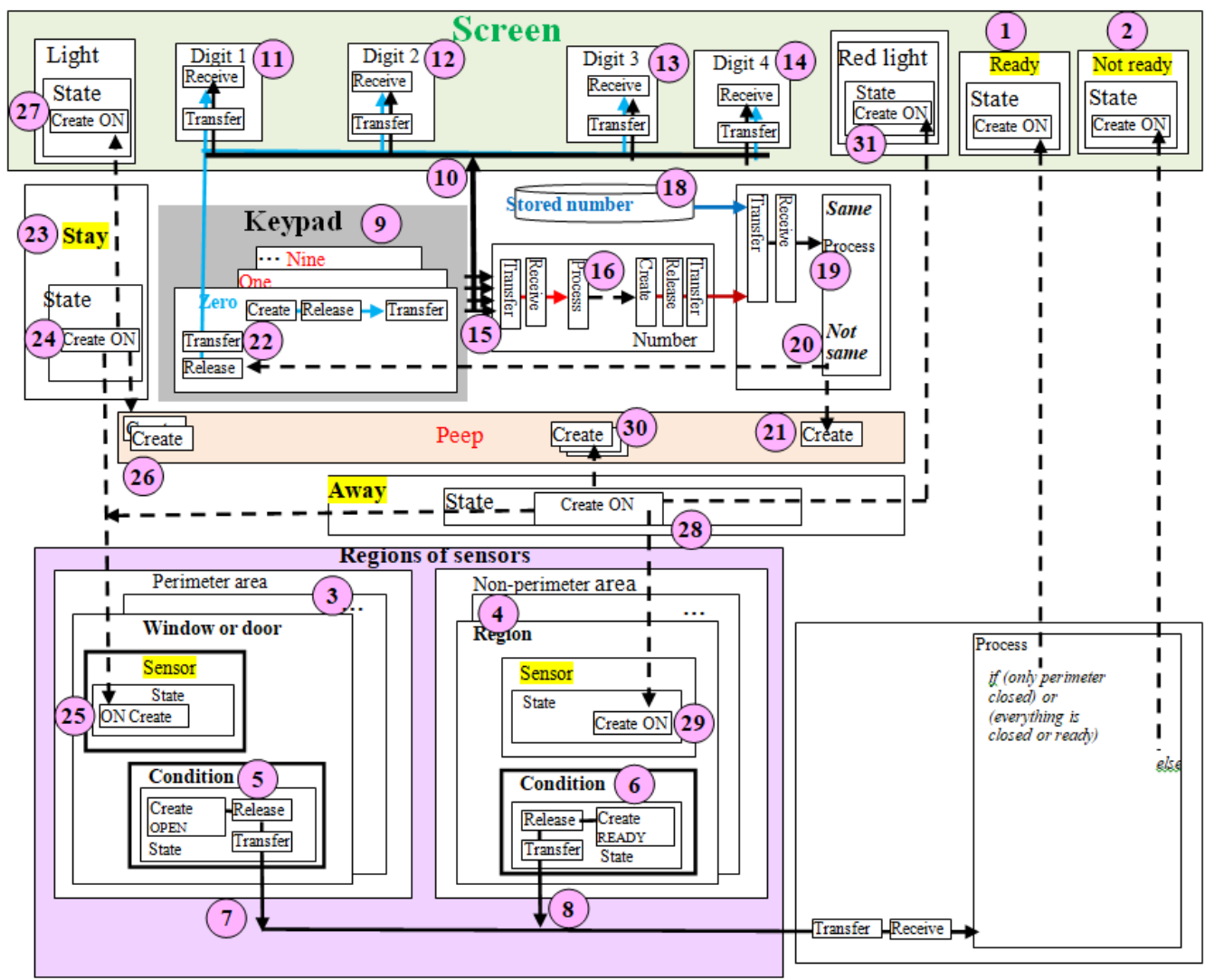

Fig. 10 The static TM model of the scenario of the homeowner using the control panel.

## 7. Dynamic Model

The static model represents only the steady (static) whole, so it is necessary to analyze the underlying decompositions, called regions, where behavior can happen (the potentiality of dynamism). The TM model fuses space and time into a single dynamic model. The static description is projected as the spatiality/actionality (region). In fact, a region is a subdiagram of the static model that includes spatial boundaries and actions.

A union of this TM spatiality/actionality with time defines events where an event blends such a spatiality/actionality thimac with time. Fig. 11 shows the event triggering (the homeowner pushes a number key) the generation of one digit.

The static model, S, represents only the steady (static) whole, so it is necessary to analyze the underlying decompositions, called regions, where behavior can happen (the potentiality of dynamism). Representing events by their regions, Fig. 12 shows the events in the scenario of the homeowner using the control panel.

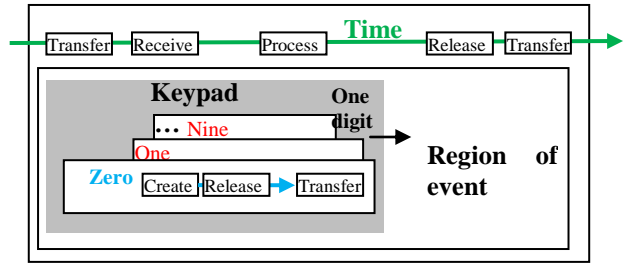

**Fig. 11 The event triggering the keypad to generate one digit.**

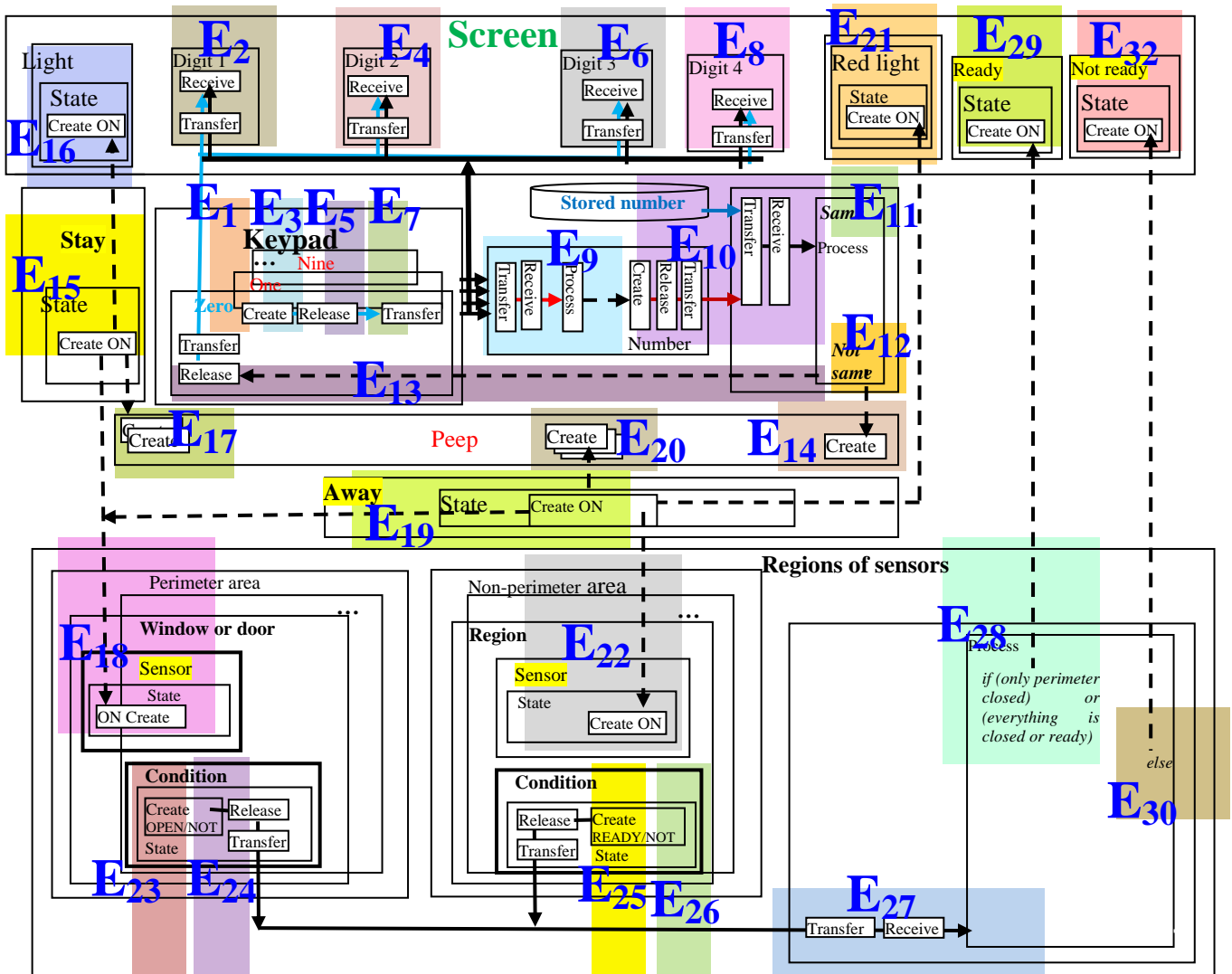

**Fig. 12 The static TM model of the scenario of the homeowner using the control panel.**

Accordingly, we select the following events.

Event 1 ($E_1$): The homeowner triggers the creation of a first digit.
Event 2 ($E_2$): The first digit is displayed on the screen.
Event 3 ($E_3$): The homeowner triggers the creation of a second digit.
Event 4 ($E_4$): The second digit is displayed on the screen.
Event 5 ($E_5$): The homeowner triggers the creation of a third digit.
Event 6 ($E_6$): The third digit is displayed on the screen.
Event 7 ($E_7$): The homeowner triggers the creation a fourth digit.
Event 8 ($E_8$): The four digits are displayed on the screen.
Event 9 ($E_9$): The four digits are converted into a number.
Event 10 ($E_{10}$): The number is compared with the stored password.
Event 11 ($E_{11}$): The two numbers are equal.
Event 12 ($E_{12}$): The two numbers are not equal.
Event 13 ($E_{13}$): Zeros flow to all digits on the screen.
Event 14 ($E_{14}$): Beeping once
Event 15 ($E_{15}$): "Stay" is selected.
Event 16 ($E_{16}$): The "stay" light is on.
Event 17 ($E_{17}$): Beeping twice
Event 18 ($E_{18}$): All sensors in the perimeter are set.
Event 19 ($E_{19}$): "Away" is selected.
Event 20 ($E_{20}$): Beeping three times
Event 21 ($E_{21}$): The red light is on.
Event 22 ($E_{22}$): All sensors are set.
Event 23 ($E_{23}$): Some doors or windows are open.
Event 24 ($E_{24}$): All doors and windows are closed.
Event 26 ($E_{26}$): All sensors in the non-perimeter area are ready.
Event ($E_{27}$): Data on the conditions of all sensor areas are processed.
Event 28 ($E_{28}$): All sensors are okay; the perimeter sensors are okay (no open doors or windows, or sensors not ready).
Event 29 ($E_{29}$): Ready state
Event 30 ($E_{30}$): Sensors in the perimeter area are not okay (e.g., a door or window is open).
Event 31 ($E_{31}$): Not-ready state

Fig. 13 shows the behavioral model of this part of the SafeHome project.

## 9. Expanding While Preserving Model Singularity

Then, Pressman and Maxim [6] introduced the class diagram, tying it to scenarios where objects (classes) are manipulated as an actor interacts with the system. They provided a sensor class (Fig. 14) for the SafeHome security function. A sensor is described in terms of attributes and operations.

Our method of contrasting a TM and UML is to apply the UML modeling step to the corresponding TM. Fig. 15 shows this expanding TM static model. Accordingly, the attributes of the sensor are simply added (e.g., name and location as shown in the orange [shaded] subdiagram and labeled with a capital "A" in Fig. 15 [circle 1]).

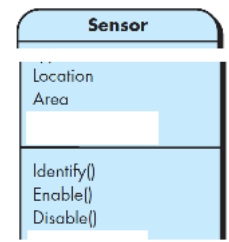

**Fig. 14 Class diagram (Partial, from [6]).**

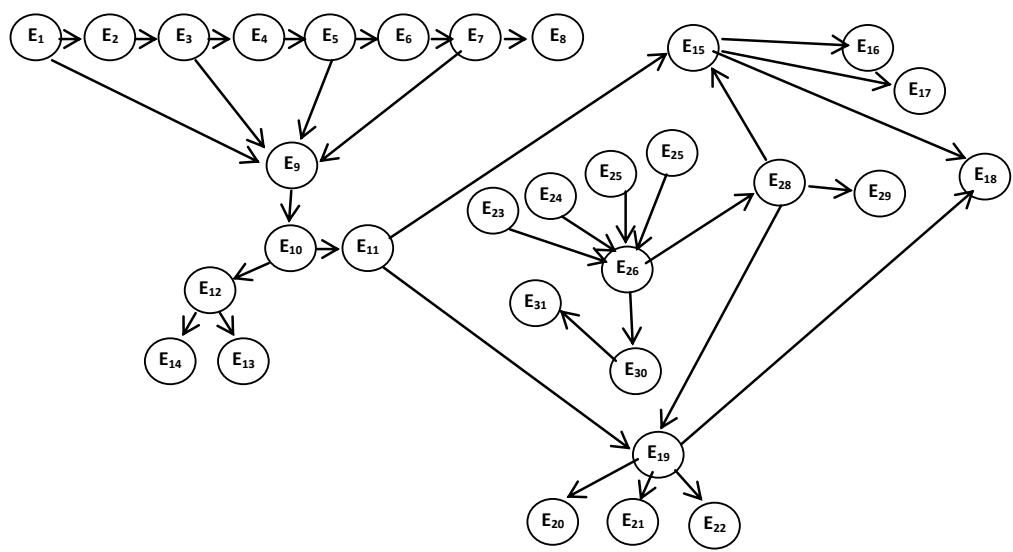

**Fig. 13 The behavioral model.**

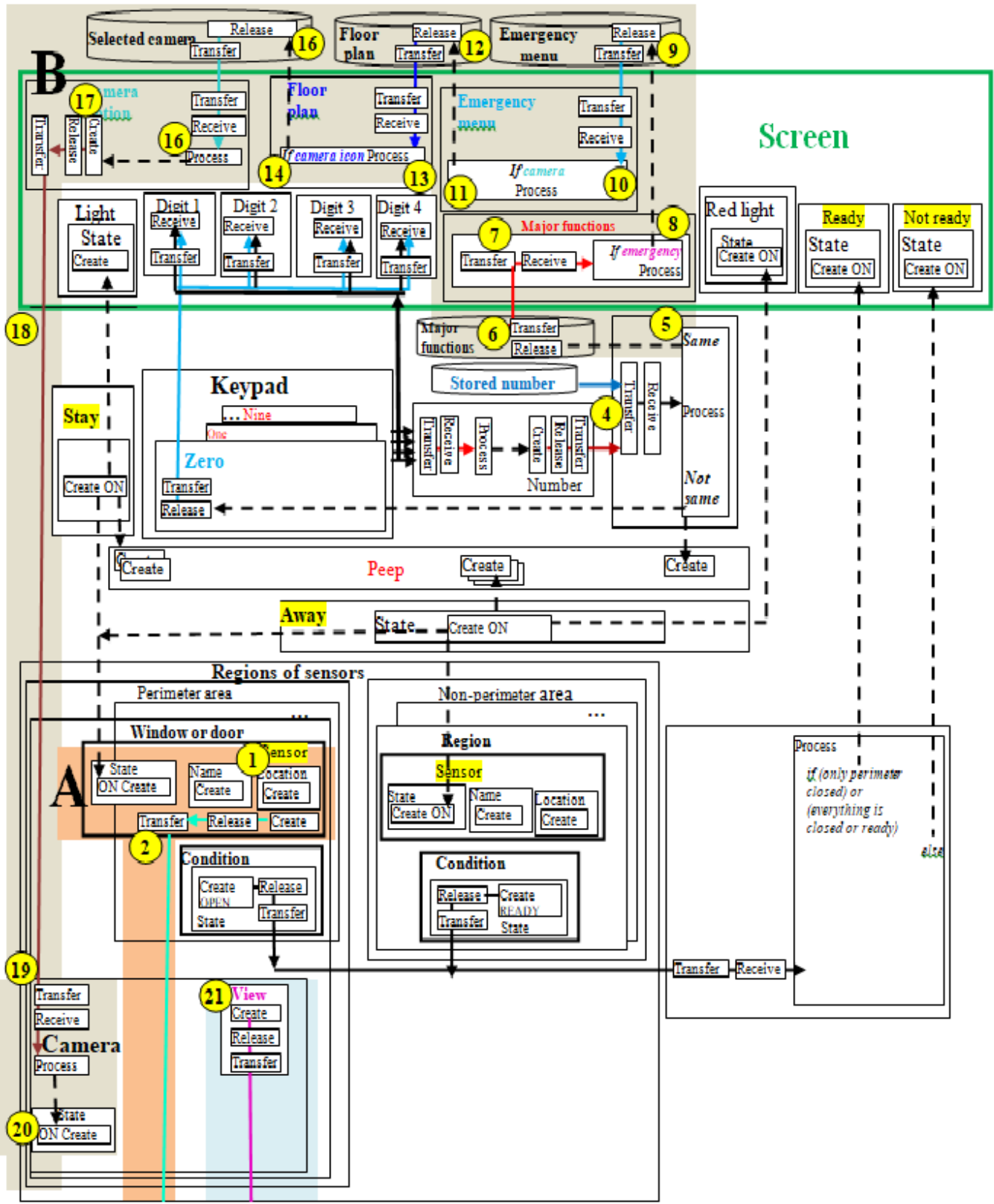

Fig. 15 Expanding the TM model of the scenario of the homeowner using the control panel.

Additionally, the operation of a sensor can be incorporated into the TM diagram in the usual way. For example, in the same area of "A," the record of a sensor flows (2) to the screen to be displayed (3). The class diagram follows the class diagram in Pressman and Maxim's [6] book. They mentioned that the requirements model must provide modeling elements that depict behavior; hence, a state diagram for "software embedded within the SafeHome control panel that is responsible for reading user input" is introduced. The state diagram is called reading commands and includes state variables.

Note that such a diagram is for "embedded software," which would completely baffle students when trying to follow the development of the SafeHome project. Accordingly, we ignore this state diagram because the issue of behavior in a TM comes after the development of the static model.

Then, Pressman and Maxim's [6] book highlights an additional piece of the SafeHome project in terms of another scenario:
1. The homeowner logs onto the SafeHome Products website.
2. The homeowner enters his or her user ID.
3. The homeowner enters a password (modified).
4. The system displays all major function buttons.
5. The homeowner selects "surveillance" from the major function buttons.
6. The homeowner selects "pick a camera."
7. The system displays the floor plan of the house.
   Etc....

This can easily be incorporated into the TM diagrams because all pieces of the SafeHome project complement one another, just like the puzzle pieces of pictures and images. UML is a stovepipe system to generate separate pieces, whereas a TM is a way to provide the total picture.

Fig. 15 also shows this additional scenario in the beige-colored subdiagram labeled "B." For simplicity's sake, we use the four digits discussed previously as the password. Thus, when the input number and the stored number (circle 4) are the same (5), this triggers the sending of the main functions (6) to be displayed on the screen (7). If the homeowner selects "emergency," this triggers the displaying of the emergency menu (9 and 10). If the homeowner selects "camera," then this triggers displaying the floor plan (11 and 12). If the homeowner selects a certain camera icon, this triggers the displaying of the available option for the camera (15 and 16). Based on the homeowner's choice, the camera is turned on (17, 18, 19, and 20—in the door and window area). This would send the view to the screen to be displayed (21 and 22).

The expanding of the static model, as mentioned above, will continue as additional information and requirements are presented.

## 10. Conclusion

This paper contributes to establishing a broad understanding of conceptual modeling instead of presenting it as an object-oriented venture using UML. The direct goal was to provide a better modeling foundation for soft engineering students. The contrast between model multiplicity and model sincerity, even as an introductory topic to UML (one chapter) in current software engineering texts, will establish a greater appreciation of the advantages and limitations of UML itself.

Of course, TM has its own advantages. It provides enough precision but is still easy enough for all stakeholders to use. Its complexity is apparent with TM's five generic actions and the repeatability of applying modeling in terms of these actions. Similarly, the difference between static representation and behavior is applied at different levels of modeling.

One benefit of the paper is the apparent suitability of the TM diagrammatic method for expressing difficult notions, such as time. Future work will involve applying the methods for other philosophical approaches to time. Many techniques can be utilized, such as "zoom in" and "zoom out." Future research could examine the degree of students' improvement (e.g., learning UML) when they are taught using the proposed approach.

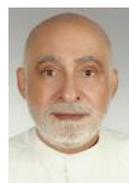

**Sabah S. Al-Fedaghi** is an associate professor in the Department of Computer Engineering at Kuwait University. He holds an MS and a PhD from the Department of Electrical Engineering and Computer Science, Northwestern University, Evanston, Illinois, and a BS from Arizona State University. He has published more than 350 journal articles and papers in conferences on software engineering, database systems, information ethics, privacy, and security. He headed the Electrical and Computer Engineering Department (1991–1994) and the Computer Engineering Department (2000–2007). He previously worked as a programmer at the Kuwait Oil Company.